\documentclass[showpacs,preprintnumbers,amsmath,amssymb,twocolumn]{revtex4}
\usepackage{graphicx}   % Include figure files
\usepackage{dcolumn}    % Align table columns on decimal point
\usepackage{bm}         % bold math
\usepackage{ifpdf}
\usepackage{graphicx,amssymb,lineno}
\usepackage{amssymb,lineno,amsfonts}
\usepackage{graphicx}   % Include figure files
\usepackage{dcolumn}    % Align table columns on decimal point
\usepackage{bm}         % bold math
\usepackage{bbm}
\usepackage{mathrsfs}
\usepackage{upgreek}
\usepackage{mathtools}

\begin{document}

\title{Initialization of NMR Quantum Registers using Long-Lived Singlet States}

\author{Soumya Singha Roy and T. S. Mahesh}
\address{NMR Research Center, \\
Indian Institute of Science Education and Research (IISER), Pune 411008, India}

\date{\today}

\begin{abstract}
{
An ensemble of nuclear spin-pairs under certain conditions is
known to exhibit singlet state life-times much longer than other 
non-equilibrium states. This property
of singlet state can be exploited in quantum information processing
for efficient initialization of quantum registers.  Here we describe a general 
method of initialization and experimentally demonstrate it with two-, 
three-, and four-qubit nuclear spin registers.
}
\end{abstract}

\maketitle

\section{Introduction}
It has been established theoretically that quantum systems
are far more capable than their classical counterparts for
certain computational tasks \cite{chuangbook}.  
Although various types of quantum
systems are being explored, the realization of a general purpose quantum 
processor remains a technological challenge \cite{laflammerev}. 
Proof-of-principle type
experimental demonstrations have been carried out for several
problems including factorization, 
unsorted database search, and quantum simulations 
using Nuclear Magnetic Resonance (NMR) \cite{coryrev04,dieterrev}.  
Nuclear spins in an ensemble of molecules 
present a convenient architecture to simulate a quantum register.
The practical demonstrations of Quantum Information Processing (QIP)
using such registers are greatly benefited by the long coherence
times of nuclear spins and the well developed control techniques of 
NMR \cite{coryrev04,dieterrev}.

In order to carry out information processing, a quantum register
must satisfy a set of criteria laid out by DiVincenzo \cite{divincenzo}.
An important criterion is the ability to precisely initialize the 
register to a desired ket of the computational basis.  Nuclear
spins in room temperature and at ordinary magnetic fields 
exist in a highly mixed state
and therefore preparing a pure state is not straightforward
\cite{goldman,Abragam,LevBook}.  
The spin temperature can nevertheless be reduced by 
using parahydrogens \cite{Anwar} or by using
Dynamic Nuclear Polarization \cite{Morley}.  In future, either or both of
these techniques may be available for preparing NMR quantum registers 
into almost pure states \cite{laflammerev}.  
The existing approach for initializing NMR registers is
however based on specially prepared mixed states known as 
`pseudopure states', which are isomorphic to pure states for 
several computational problems \cite{corypps,chuangpps}.

Consider an ensemble of identical molecules each having $n$ spin-1/2 nuclei in a magnetic
field.  The Zeeman Hamiltonian ${\cal H}_z = h \sum_j \nu_0^j I_z^j$,
is characterized by the frequency $\nu_0^j$ of Larmor precession,
and the z-component of the spin angular momentum operator $I_z^j$ of spins $j = 1 \cdots n$
\cite{LevBook}.
In NMR-QIP, the eigenstates $\vert \pm 1/2 \rangle$ of ${\cal H}_z^j$
are labeled as $\vert 0 \rangle$ and $\vert 1 \rangle$ states
of a qubit, and the  multi-spin eigenbasis 
$\{\vert 00 \cdots 00 \rangle, \vert 00 \cdots 01 \rangle, \cdots \}$
is treated as the computational basis.
In thermal equilibrium at ordinary room temperature $T$, 
the Boltzman factor $kT$ is much larger than
the Zeeman energy gaps, so that the density matrix can be expanded as
\begin{eqnarray}
\rho_\mathrm{eq} = 2^{-n}e^{-{\cal H}_z/{kT}} \approx 2^{-n}({\mathbbm 1} + \rho_\Delta).
\label{rhoeq}
\end{eqnarray}
%where ${\mathbbm 1}$ forms the 
Here identity ${\mathbbm 1}$ corresponds to a background of uniformly populated levels 
and the trace-less part
$\rho_\Delta = \sum_j \epsilon_j I_z^j$ is known as the deviation density matrix.
The dimensionless numbers $\epsilon_j = -h \nu_0^j / kT$ 
have magnitudes $\sim 10^{-4}$ for protons in 
currently available magnetic fields
at ordinary room temperatures.  The deviation density matrix
represents the unequal population distribution (over the uniform background)
which lead to the observable magnetizations.  Thus preparing a pure ground state i.e.,
$\vert 0 0 0 \cdots 0 \rangle$ in an NMR quantum register is rather difficult.  
Nevertheless, Cory et al \cite{corypps} and Chuang et al \cite{chuangpps}
have independently pointed out that often a pure state $\vert \psi \rangle \langle \psi \vert$ can 
be simulated by the pseudopure state
\begin{eqnarray}
\rho_\mathrm{pps} \approx 2^{-n}\left[ (1-\epsilon'){\mathbbm 1} + 
2^n \epsilon' \vert \psi \rangle \langle \psi \vert \right].
\label{rhopps}
\end{eqnarray}
Here $\epsilon'$ is a measure of the magnetization retained in
the pseudopure state and it usually gets halved with every additional 
qubit \cite{MaheshBEN}.
The unit background is invariant under the Hamiltonian
evolution, does not lead to NMR signal and is ignored \cite{corypps}.
Thus the equilibrium density matrix of a single spin-1/2 nucleus
is always in a pseudopure state.  Initializing a multi-spin system into a pseudopure state
however is essentially a non unitary process \cite{dieterrev}.  
Several methods have earlier
been proposed and they involved averaging 
of magnetization modes over the sample space 
(called `spatial averaging' \cite{corysp}) or 
over spin space (called `logical labeling' \cite{chuangpps,kavita}) or over
several transients (called `temporal averaging' \cite{knillpps}).
In some cases subsystem pseudopure states are easier to prepare 
either by transition selective pulses \cite{maheshpps}
or by coherence selection using pulsed field gradients \cite{benchmark},
but these methods invariably result in loss of a qubit for further computation.

In the next section we propose a different approach
that exploits long life-times of certain special states called
`singlet states'. 
The following section details the
experimental demonstrations on model systems consisting of two, three 
and four-qubit NMR registers.

\section{Singlet States and Register Initialization}
\subsection{Singlet States}
\label{secsingstates}
The Hamiltonian for an ensemble of spin-1/2 nuclear pairs of same isotope, in 
the RF interaction frame, can be expressed as
\begin{eqnarray}
{\cal H^{\mathrm{eff}}} =  h \left[ 
         \frac{\Delta\nu}{2}  I_z^1 
         - \frac{\Delta\nu}{2}I_z^2 
         +  J I^1 \cdot I^2
         +  \nu_{12} I_x^{1,2}
          \right].
\label{heff}         
\end{eqnarray}
Here the RF frequency is assumed to be at the
mean of the two Larmor frequencies, and
$\Delta \nu$, $J$ and $\nu_{12}$ correspond  respectively to the
difference in Larmor frequencies (chemical shift difference), the
scalar coupling constant and the RF amplitude (all in Hz).  
In the limiting case of $\Delta \nu \rightarrow 0$, the system is
said to have magnetic equivalence, and the singlet state 
$\vert S_0 \rangle = (\vert 0 1 \rangle - \vert 1 0 \rangle)/\sqrt{2}$, and
the triplet states $\vert T_1 \rangle = \vert 0 0 \rangle$, 
$\vert T_0 \rangle = (\vert 0 1 \rangle + \vert 1 0 \rangle)/\sqrt{2}$,
and $\vert T_{-1} \rangle = \vert 1 1 \rangle$ 
form an orthonormal eigenbasis of the internal Hamiltonian 
${\cal H}_{\mathrm{eq}}^{\mathrm{eff}} = h J I^1 \cdot I^2$
\cite{LevBook}.  

The non-equilibrium nuclear spin states decay toward equilibrium state (\ref{rhoeq})
with a time constant called `spin-lattice' relaxation time
constant $T_1$ \cite{Abragam}.  Since most of the NMR experiments involve preparation and 
detection of non-equilibrium spin states, it was generally accepted that the
duration of any NMR transient is limited by $T_1$, although the theoretical
limit of transverse relaxation is $2T_1$ \cite{LevBook}.
Levitt and co-workers demonstrated that under the Hamiltonian 
${\cal H}_{\mathrm{eq}}^{\mathrm{eff}}$
the decay constant of singlet state $\vert S_0 \rangle$ is much larger than $T_1$, 
and hence called it a long-lived state \cite{LevPRL04,LevittJACS04}.  
The phenomenon is akin to the decoherence-free-subspace (DFS)
which is well known in QIP \cite{LidarDFS}.
Detailed theoretical analyses of singlet state decay have  been
provided by Levitt and co-workers \cite{LevittJCP05,LevJCP09} and by 
Karthik and Bodenhausen \cite{BodenJMR06}.
The long life times of singlet states have been attributed to the fact 
that the singlet states are immune to the dominant intramolecular dipole-dipole 
relaxation mechanism, which is symmetric with respect to the exchange of spin states.

Experimentally, singlet decay constants upto $36 T_1$ 
have been reported \cite{sarkar36}, and in another instance singlet lived
till about half an hour \cite{LevittJACS08}.
Overcoming the $T_1$ limit has
motivated a number of novel applications including studying slow molecular dynamics and transport 
processes \cite{sarkar36,BodenJACS05}, 
precise measurements of NMR interactions \cite{LevPRL09}, the 
transport and storage of hyper polarized nuclear spin order 
\cite{GrantJMR08,BargonJCP06,WeitekampJACS08,BodenPNAS09,AdamsScience09,WarrenScience09},
homogeneous line narrowing in spectroscopy \cite{bodenprl10}, and
determining molecular torsion angles \cite{LevJACS10}.
Recently we had reported that high-fidelity singlet states
can be easily prepared and characterized in two-spin 
systems under various experimental conditions \cite{maheshjmr10}.  
In the following we describe the preparation and detection of singlet states and
then elucidate the initialization of quantum registers.

\subsection{Preparation and detection of singlet states}
\label{secsingprepdet}
The state $\vert S_0 \rangle \langle S_0 \vert -\vert T_0 \rangle \langle T_0 \vert$
can be easily prepared from the equilibrium density matrix $I_z^1 + I_z^2$ 
by using the propagator
\begin{eqnarray}
U_1^{1,2} = 
      \mathrm{e}^{-i \frac{\pi}{4} (I_z^1 - I_z^2)}
 \mathrm{e}^{-i \frac{\pi}{2} I_y^{1,2} }
 \mathrm{e}^{-i \frac{\pi}{2} (I_z^1 - I_z^2 )}
 \mathrm{e}^{-i \pi I_z^1 I_z^2}
 \mathrm{e}^{-i \frac{\pi}{2} I_x^{1,2}}.
\label{singprep}
\end{eqnarray} 
With certain approximations the above propagator can be constructed from the 
Hamiltonians similar to the one in (\ref{heff}) \cite{LevJMR06}.
The equivalence Hamiltonian ${\cal H}_{\mathrm{eq}}^{\mathrm{eff}}$ may be realized
by suppressing the chemical shift $\Delta \nu$ 
either by vanishing the Zeeman field \cite{LevPRL04},
or by using a `spin-lock' \cite{LevittJACS04,BodenCPC07}.  
In this work we employ the latter technique.  The spin-lock
may be achieved by applying a long low-power unmodulated RF 
\cite{LevittJACS04}, or by
specially designed phase modulated sequence which were
originally used for spin-decoupling \cite{BodenCPC07}.

The singlet states by themselves are inaccessible to
macroscopic observables, but
can be indirectly detected by removing the equivalence and
transforming to observable single quantum coherence using the propagator
\cite{LevPRL04,LevittJACS04}
\begin{eqnarray}
U_D^{1,2} = 
\mathrm{e}^{-i \frac{\pi}{2} I_x^{1,2}}
\cdot \mathrm{e}^{-i \frac{\pi}{4} (I_z^1 - I_z^2)}.
\label{singdet}
\end{eqnarray}   
Alternatively, a more detailed and quantitative analysis of singlet states 
may be carried out  using density matrix tomography \cite{maheshjmr10}. 

\subsection{Initializing NMR Registers}
\label{secsingini}
\begin{center}
\begin{figure}
\hspace*{-.3cm}
\includegraphics[width=9cm]{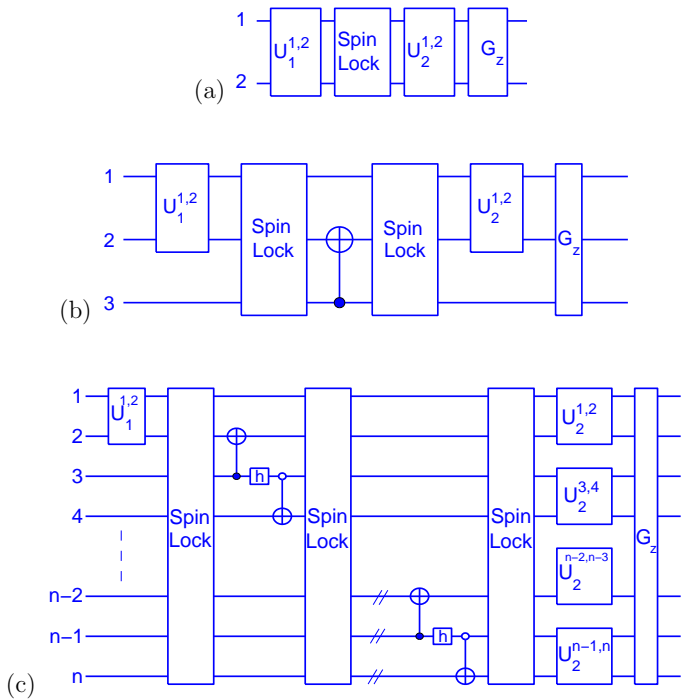} 
\caption{
The circuit diagrams for initializing (a) 2-qubit, (b) 3-qubit, and
(c) n-qubit registers.
In (c), the cNOT gates with open circles correspond to NOT operation if the
control is 0 and identity if the control is 1. The h-gate corresponds to 
pseudo-Hadamard: 
$\vert 0 \rangle \stackrel{\mathrm{h}}{\rightarrow} (\vert 0 \rangle - \vert 1 \rangle)/\sqrt{2}$
and 
$\vert 1 \rangle \stackrel{\mathrm{h}}{\rightarrow} (\vert 0 \rangle + \vert 1 \rangle)/\sqrt{2}$.
}
\label{q123pps} 
\end{figure}
\end{center}

The circuit for initializing a 2-qubit NMR register 
via singlet states is shown in Fig.\ref{q123pps}a.
An initially imperfect 
singlet density matrix gets purified during the spin-lock period 
as a result of the long life time,
while the 
artifact coherences are destroyed by relaxation process as well as 
the inhomogeneities in the spin-lock itself.  There exist
optimal spin-lock conditions at which one obtains singlet states
with high fidelity \cite{maheshjmr10}.
Once the singlet state is prepared with high fidelity, the conversion 
$\vert S_0 \rangle \rightarrow \vert 01 \rangle$ 
can be easily achieved by the propagator
\begin{eqnarray}
U_2^{1,2} = 
      \mathrm{e}^{i \frac{\pi}{2} I_x^{1,2}}
\cdot \mathrm{e}^{-i \pi I_z^1 I_z^2}
\cdot \mathrm{e}^{-i \frac{\pi}{2} I_x^{1,2}}.
\label{singtopps}
\end{eqnarray}
Finally a pulsed field gradient $G_z$ can be used to destroy the
residual single and multiple quantum coherences.
If necessary, other pseudopure states can be obtained simply by 
applying NOT gates.

This procedure of initialization can be extended to multiqubit systems,
since it is known
that a spin-pair may exhibit long-lived singlet state if it
is situated sufficiently away from other spins \cite{LevJACS10}.  
The procedure of a 3-qubit initialization is shown in Fig.\ref{q123pps}b.
We first prepare a two-qubit
singlet and apply a cNOT gate on qubit-2 (controlled by qubit-3).  
Subsequent spin-lock and $U_2^{1,2}$
gate on qubits 1 and 2 initializes a three qubit system 
into $\vert 010 \rangle$ state.  
The circuit can be understood as follows.  After 
preparing the singlet on qubits 1 and 2, the third qubit remains in a mixed
state with a probability $p_0$ of being in state $\vert 0 \rangle$ and
a probability $p_1$ of being in state $\vert 1 \rangle$.
The cNOT gate transforms the mixed state according to:
\begin{eqnarray}
\vert S_0 \rangle \langle S_0 \vert \otimes 
\left( p_0 \vert 0 \rangle \langle 0 \vert +  
 p_1 \vert 1 \rangle \langle 1 \vert \right)
\stackrel{\mathrm{cNOT}}{\longrightarrow}  \nonumber \\
p_0 \vert S_0 \rangle \langle S_0 \vert \otimes 
\vert 0 \rangle \langle 0 \vert +  
p_1 \vert \phi^- \rangle \langle \phi^- \vert \otimes 
\vert 1 \rangle \langle 1 \vert.
\label{pps3qexp}
\end{eqnarray}
First term in the output is singlet on qubits 1 and 2 and therefore survives during
the second spin-lock where as the other term consisting of 
the Bell state $\vert \phi^- \rangle = (\vert 00 \rangle - \vert 11 \rangle)/\sqrt{2}$
decays fast.  The singlet
is ultimately transformed into $\vert 01 \rangle$
by $U_2^{1,2}$ and thus we obtain $\vert 010 \rangle$ pseudopure state with 
a good approximation.  

A general scheme for the initialization of an n-qubit NMR quantum register
is described in Fig.\ref{q123pps}c.  The circuit can be analyzed in a similar
way as in the 3-qubit case.  The fact that only nearest-neighbor 
interactions are used is highly advantageous in practice.  
Experimentally, the spin-lock of multiple singlet pairs can be achieved 
using sophisticated modulated RF sequences as described in the next section.

\section{Experiments}
The following experiments are carried out in Bruker 500 MHz 
NMR spectrometer at an ambient temperature of 300 K.  
We used strongly modulated pulses for designing high fidelity
local gates as well as cNOT gates  \cite{fortunato, MaheshNGE}.
The spin-lock was achieved by WALTZ-16 - a phase modulated RF sequence,
which is routinely used in broadband spin decoupling \cite{maheshjmr10}.  
Spectra corresponding to pseudopure states are obtained by linear detection
scheme using small flip angle RF pulses.  Since
the pseudopure states have one energy level more populated than
the equal distribution in all others, 
the spectrum should consist ideally of only one transition 
per qubit in each case. Quantitative analyses of the pseudopure states
are carried out using extended versions of density matrix tomography 
described in reference \cite{maheshjmr10}.  Finally, the success of 
the experimental state $\rho$ in achieving a target pseudopure state
$\rho_\mathrm{pps}$ is measured by calculating the correlation \cite{fortunato},
\begin{eqnarray}
\langle \rho_\mathrm{pps} \rangle = \frac{\mathrm{trace}\left[ \rho \cdot \rho_\mathrm{pps} \right]}{\sqrt{ \mathrm{trace} \left[\rho^2 \right] \cdot  \mathrm{trace} \left[\rho_\mathrm{pps}^2 \right]}}.
\label{corr}
\end{eqnarray}
Often only the diagonal elements of the density matrices are 
relevant and in such cases, the `diagonal correlation' can be expressed
by replacing all the operators in the above expression by their
diagonal parts.
In the following we describe the individual cases of two-, three- and four-qubit
registers.

\subsection{2-qubit register}
\begin{center}
\begin{figure}
\includegraphics[width=7.2cm]{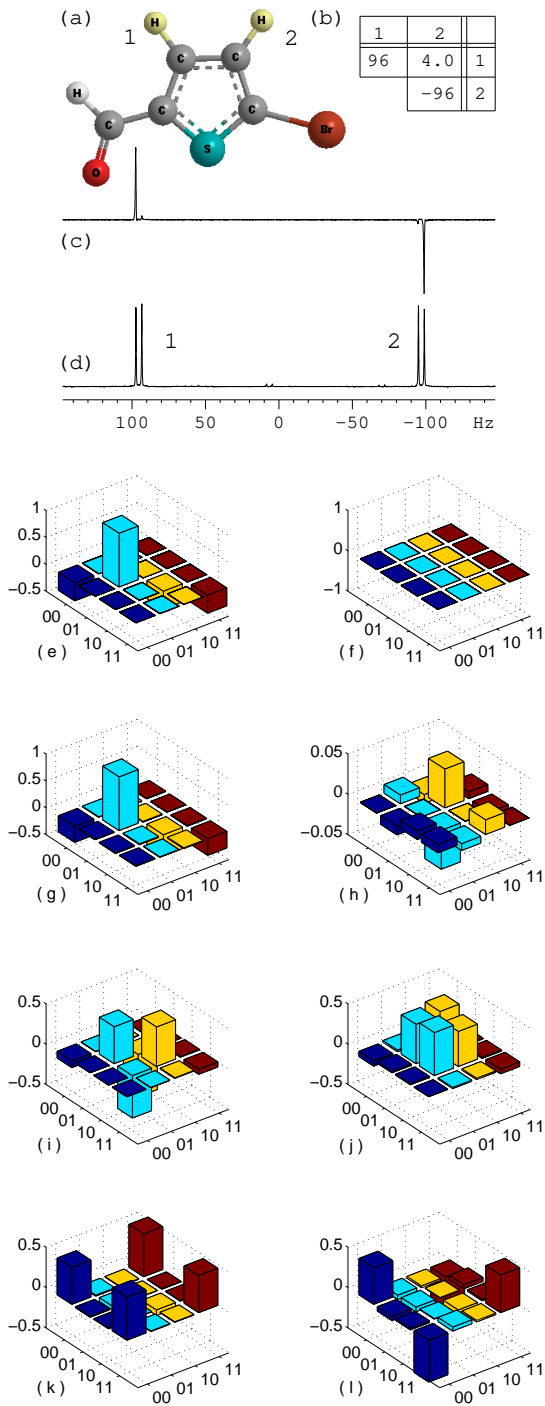}
\vspace*{0.1cm}
\caption{
The molecular structure (a) and Hamiltonian parameters (b) of 
5 -bromothiophene-2-carbaldehyde dissolved in CD$_3$OD, forming
a homonuclear two-qubit register.
In (b) diagonal and off-diagonal elements correspond to the chemical shifts and
the scalar coupling constant respectively (in Hz).
The $^1$H spectra correspond to
the pseudopure state (c) and the equilibrium mixed state (d).
The barplots (e-h) correspond to the real (e,g) and imaginary (f,h)
parts of theoretical (e,f) and experimental (g,h) pseudopure
$\vert 01 \rangle$ state.  The remaining barplots
correspond to the experimental real parts of
$\vert S_0 \rangle$ (i),
$\vert \psi^+ \rangle$ (j),
$\vert \phi^+ \rangle$ (k), and
$\vert \phi^- \rangle$ (l).
}
\label{pps2q}
\end{figure}
\end{center}

The two-qubit system, Hamiltonian parameters, and the corresponding
pseudopure and the reference spectra are shown in Fig.\ref{pps2q}(a-d).
As shown in Fig.\ref{q123pps}a, the experiment involved preparing singlet
using $U_1^{1,2}$, followed by RF spin-lock with amplitude 2 kHz and
duration 12.4 s, which are optimized for high singlet content \cite{maheshjmr10}.
The decay constant for singlet state was 16.2 s approximately three
times the $T_1$ values of the two spins.
The singlet is then converted into $\vert 01 \rangle$ 
pseudopure state using $U_2^{1,2}$.  A final gradient pulse served
to destroy the artifact coherences.  
The bar plots showing the real and imaginary parts of the theoretical
and experimental density matrix are shown in Fig.\ref{pps2q}(e-h).  
A very high correlation of 0.995 is obtained with $\vert 01 \rangle$
pseudopure state.  

For some quantum algorithms a good starting point
may be the singlet state itself, which can be extracted
directly after the spin-lock shown in Fig.\ref{q123pps}a.
The real part of the experimental singlet density matrix
shown in Fig.\ref{pps2q}i has a correlation of 0.991. 
If necessary, initialization to other Bell states, 
starting from the singlet state, can be carried out easily:
\begin{eqnarray}
\vert S_0 \rangle & \stackrel{\mathrm{e}^{i\pi I_z^1}}{\longrightarrow}    &  \vert \psi^+ \rangle = (\vert 01 \rangle + \vert 10 \rangle)/\sqrt{2}, \nonumber \\
\vert S_0 \rangle & \stackrel{\mathrm{e}^{i\pi I_x^1} \cdot \mathrm{e}^{i\pi I_z^1}}{\longrightarrow} & \vert \phi^+ \rangle = (\vert 00 \rangle + \vert 11 \rangle)/\sqrt{2}, \nonumber \\
\vert S_0 \rangle & \stackrel{\mathrm{e}^{i\pi I_x^1}}{\longrightarrow}               & \vert \phi^- \rangle = (\vert 00 \rangle - \vert 11 \rangle)/\sqrt{2}.
\end{eqnarray}
The z-rotation in the above propagators can be implemented by simply
chemical shift evolution for a period $1/(2\Delta \nu)$, and the 
qubit selective x-rotation can be
implemented by using a strongly modulated pulse.
The experimental density matrices corresponding to
these Bell-states have respective correlations 
0.987 (Fig.\ref{pps2q}j),
0.982 (Fig.\ref{pps2q}k), and
0.968 (Fig.\ref{pps2q}l).

\subsection{3-qubit register}
\begin{center}
\begin{figure}
\includegraphics[width=8.2cm]{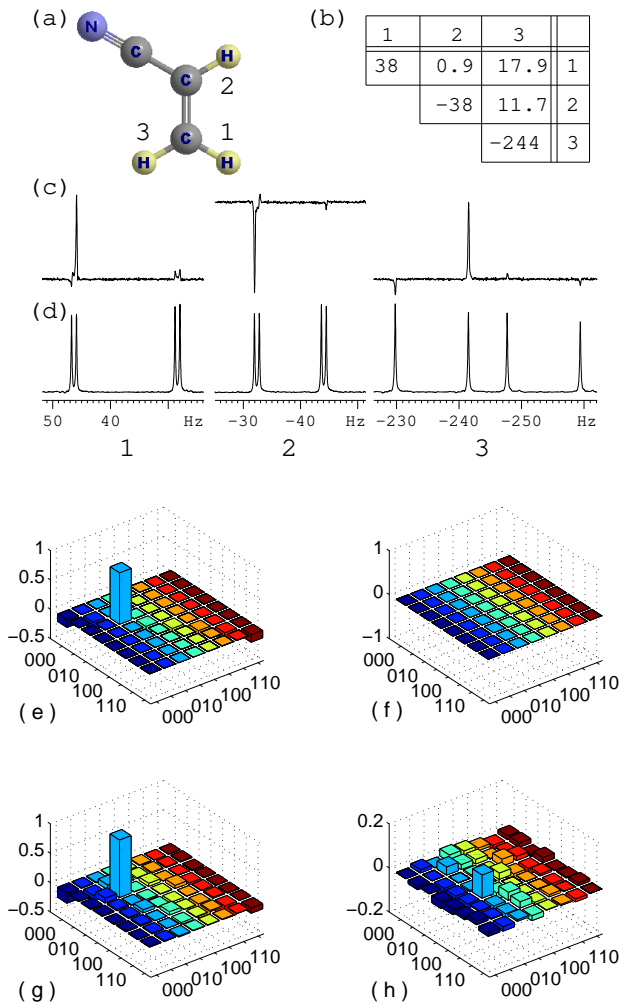}
\caption{
The molecular structure (a) and Hamiltonian parameters (b) of 
acrylonitrile dissolved in CDCl$_3$, forming a homonuclear
3-qubit register.
In (b) diagonal elements are chemical shifts (in Hz) and the off-diagonal
elements are the scalar coupling constants (in Hz).  
The $^1$H spectra correspond to the pseudopure state (c) 
and the equilibrium mixed state (d). 
The bar plots are showing the real (e,g) and imaginary (f,h)
parts of theoretical (e,f) and experimental (g,h) pseudopure
$\vert 010 \rangle \langle 010 \vert$ state.  
}
\label{pps3q}
\end{figure}
\end{center}

The three-qubit system, Hamiltonian parameters and the corresponding
pseudopure and the reference spectra are shown in Fig.\ref{pps3q}(a-d).
The decay constant for singlet state of spins 1 and 2 was about 18 s, 
approximately three times of their $T_1$ values.
The pseudopure state was prepared using the circuit shown in 
Fig.\ref{q123pps}b.
Each of the two spin-locks were achieved using 6.3 s of 500 Hz WALTZ-16 
modulations.  The cNOT gate was implemented using a 14 segment
strongly modulated RF pulse of duration approximately 60 ms and
of fidelity 0.96.
The bar plots showing the real and imaginary parts of the theoretical
and experimental density matrix are shown in Fig.\ref{pps3q}(e-h).  
%The 3-spin tomography is an extension of the technique described in
%reference \cite{Mahesh} and will be described elsewhere.
The correlation of the experimental density matrix with the
theoretical pseudopure state is $\vert 010 \rangle$ is 0.952.  
The correlation is smaller compared to the two-qubit case, mainly
due to the errors in the cNOT gate.  Nevertheless, the diagonal
correlation is as high as 0.983.

\subsection{4-qubit register}
\begin{center}
\begin{figure}
\hspace*{-0.3cm}
\includegraphics[width=7.2cm,angle=-90]{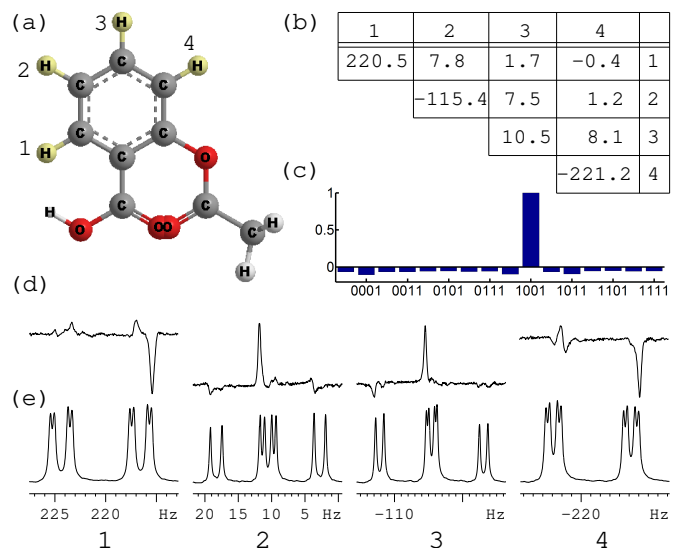}
\caption{
The molecular structure (a) and Hamiltonian parameters (b) of 
aspirin dissolved in CD$_3$OD, forming a homonuclear 4-qubit register.
In (b) diagonal elements are chemical shifts (in Hz) and the off-diagonal
elements are the scalar coupling constants (in Hz).  
The barplot in (c) displays the diagonal elements of the density matrix
obtained by tomography of the pseudopure $\vert 1001 \rangle$ state.
The $^1$H spectra correspond to the pseudopure  state (d) 
and the equilibrium mixed state (e). 
}
\label{pps4q}
\end{figure}
\end{center}

The four-qubit system, Hamiltonian parameters and the corresponding
pseudopure and the reference spectra are shown in Fig.\ref{pps4q}.
The pseudopure state was prepared using the circuit shown in 
Fig.\ref{q123pps}c.  
The singlet decay constants were about 6 s, approximately twice
the $T_1$ values of the individual spins.
We were able to carry out simultaneous 
spin-lock of two singlet pairs and initialize
a four-qubit register.  
The two spin-locks were achieved by 2 kHz WALTZ-16 modulations of
durations 2 s and 4.5 s each.
The two cNOT gates were made of 20 segments, approximately 61 ms duration
and of fidelities about 0.94.
The 10 segment h-gate was about 8.2 s long and of fidelity 0.98.
Complete tomography of a 4-qubit density matrix
is a laborious task.  After the preparation of the pseudopure state,
the non-zero quantum off-diagonal elements are efficiently
destroyed by the final gradient pulse.
Since only the diagonal elements are of main interest, we have 
carried out the four-qubit diagonal tomography \cite{MaheshNGE}.  
The bar plot showing the diagonal part of the 
experimental density matrix is shown in Fig.\ref{pps4q}c.  
The diagonal correlation is estimated to be approximately
$0.97 \pm 0.01$ with $\vert 1001 \rangle$ pseudopure state.
The first pair has collapsed to $\vert 10 \rangle$ state instead 
of $\vert 01 \rangle$, due to an additional 180 degree pulse
that was applied on qubits 1 and 2 
for refocusing purposes during $U_2^{3,4}$.

\section{Conclusions}
We have demonstrated that robust initialization of NMR quantum
registers are possible via long-lived singlet states.  
It is hard to initialize proton-based NMR registers using
traditional methods.  As a result many
popular NMR registers were based on carbon spins
or a combination of protons and carbons 
which permitted initialization by traditional methods.
The proposed method of using long-lived singlet states enables us
to initialize larger registers even though the long range
interactions are weak.
Molecules are of interest where in the
inter pair dipolar couplings are sufficiently weak to keep the
singlet states long-lived, while the covalant bond mediated scalar 
interactions among the nearest neighbor spins are sufficiently strong.  
Since para hydrogens naturally exist in singlet states, the method can be
applied directly to the initialization of registers based on parahydrogens.
While the register initialization is the first application of long-lived
singlet states for QIP, more applications, like for example 
enhancing the memory of registers, may be realized in future.
Similar techniques may be used for multi-qubit initialization
in non-NMR systems exhibiting long-lived singlet states. 

\acknowledgments
Authors gratefully acknowledge discussions with Prof. Anil Kumar 
and with Prof. G. S. Agarwal.  The use of
500 MHz NMR spectrometer at NMR Research Center, IISER-Pune
is also acknowledged.

\end{document}